\documentclass[epj]{webofc}
\usepackage[varg]{txfonts}   
\usepackage{graphicx,color} 
\usepackage{bm}       
\usepackage{amsmath}  
\usepackage{amssymb}
\usepackage{tensor}
\woctitle{21st International Conference on Few-Body Problems in Physics}
\begin{document}

\title{\boldmath \nuclide[6][\Lambda\Lambda]{He} in cluster effective field theory}

\author{
S.-I. Ando\inst{1}\fnsep\thanks{\email{sando@sunmoon.ac.kr}} \and
Y. Oh\inst{2}\fnsep\thanks{\email{yohphy@knu.ac.kr}} 
}

\institute{
Department of Information Display, Sunmoon University, Asan, Chungnam 31460, Korea
\and
Department of Physics, Kyungpook National University, Daegu 41566, Korea
}

\abstract{
The bound state of \nuclide[6][\Lambda\Lambda]{He} is studied as a three-body 
($\Lambda\Lambda\alpha$) system in a cluster effective field theory at leading order. 
We find that the system exhibits the limit cycle that is associated with the formation of bound states
called the Efimov states.
This implies that the three-body contact interaction should be promoted to the leading order.
The relationship of the binding energy and the $\Lambda\Lambda$ scattering length is discussed as well
as the role of the scale of this system. 
}

\maketitle

\section{Introduction}
\label{intro}

Although the first observation of \nuclide[6][\Lambda\Lambda]{He} was reported in 1960s~\cite{Prowse66},
its clear track was only recently caught in the emulsion experiment of KEK-E373 
Collaboration~\cite{TAAA01}, which is known as the ``NAGARA'' event.
When combined with the recent information, the two-$\Lambda$ separation energy 
$B_{\Lambda\Lambda}$ of \nuclide[6][\Lambda\Lambda]{He} is estimated to be
$B_{\Lambda\Lambda}=6.93\pm 0.16$~MeV~\cite{Nakazawa10}.
Theoretical studies on double-$\Lambda$ hypernuclei are expected to shed light on the
understanding of baryon-baryon interactions in the strangeness sector, which may help search for new 
exotic systems~\cite{HMRY10,Gal10}.

In this contribution, we present our recent work~\cite{AO14} in which the effective field theory (EFT) 
method at leading order (LO) is applied to explore the structure of \nuclide[6][\Lambda\Lambda]{He} 
as a three-body ($\Lambda\Lambda\alpha$) cluster system.%
\footnote{
This approach was also applied to relate the formation of
the bound state $\nuclide[4][\Lambda\Lambda]{H}$ and the the $S$-wave scattering of $\Lambda +
\nuclide[3][\Lambda]{H}$ below the hypertriton breakup threshold by treating 
$\nuclide[4][\Lambda\Lambda]{H}$ as a three-body ($\Lambda \Lambda d$) system 
in an EFT at LO~\cite{AYO13}.
}
One advantage of EFTs at very low energies is that it can provide a
model-independent and systematic perturbative method to analyze a system, where one can introduce 
a high momentum separation scale $\Lambda_H$ between relevant degrees of freedom of low energy 
and irrelevant degrees of freedom of high energy of the system.
For a review of this approach, we refer to Refs.~\cite{BK02,BH04} and references therein.
In the study of three-body systems, it was suggested by Bedaque \textit{et al.\/} to modify the counting 
rules so that the three-body contact interaction should be promoted to LO to renormalize the limit 
cycle~\cite{BHV99}.
The limit cycle, a cyclic singularity, in a renormalization group analysis was suggested by
Wilson~\cite{Wilson71}. 
In the unitary limit, the limit cycle is known to be associated with the formation of bound states
called the Efimov states~\cite{Efimov71}.
Whether the system exhibits the limit cycle or not is easily examined by using the homogeneous part of 
the integral equation in the asymptotic limit, which was obtained earlier by Danilov~\cite{Danilov61}.
This allows one to investigate the nature of bound states in three-body systems by using the properties 
of the limit cycle.

In the present study, we treat the $\alpha$ particle as an elementary field.
The binding energy of the $\alpha$ particle is known to be $B_4 \simeq 28.3$~MeV and 
its excitation energy is $E_1 \simeq 20.0$~MeV.
Thus the large momentum scale of the $\alpha$-cluster theory is estimated to be
$\Lambda_H \sim \sqrt{2\mu E_1}\simeq 170$~MeV with $\mu$ being the reduced mass of the 
($\nuclide[3][]{H}, \nuclide{p})$ or $(\nuclide[3][]{He},\nuclide{n})$ system so that
$\mu \simeq \frac34 m_N^{}$ with $m_N^{}$ being the nucleon mass.
On the other hand, we choose the binding momentum of $\nuclide[5][\Lambda]{He}$ as the typical 
momentum scale $Q$ of the theory.
Since the $\Lambda$ separation energy of $\nuclide[5][\Lambda]{He}$ is $B_\Lambda \simeq 3.12$~MeV, 
the binding momentum of $\nuclide[5][\Lambda]{He}$ as the $\Lambda\alpha$ cluster system is
then $\gamma_{\Lambda\alpha}^{} = \sqrt{2\mu_{\Lambda\alpha}^{} B_\Lambda}$,
where $\mu_{\Lambda\alpha}^{}$ is the reduced mass of the $\Lambda\alpha$
system.
This leads to $\gamma_{\Lambda\alpha}^{} \simeq 73.2$~MeV and our expansion parameter becomes
$Q/\Lambda_H \sim \gamma_{\Lambda\alpha}^{}/\Lambda_H \simeq 0.43$.

\section{\boldmath $\Lambda\Lambda\alpha$ bound state}
\label{sec-1}

The effective Lagrangian at LO for this system can be found in Ref.~\cite{AO14}.
With the given Lagrangian the bound state of the $\Lambda\Lambda\alpha$ system is examined through 
the homogeneous part of the coupled integral equations that are written in terms of the scattering 
amplitude of the $S$-wave $\Lambda + \nuclide[5][\Lambda]{He}$ scattering.
The obtained LO coupled integral equations are determined by three parameters:
$\gamma_{\Lambda\alpha}^{}$, $a_{\Lambda\Lambda}^{}$, and $g(\Lambda_c)$. 
Here, $\gamma_{\Lambda\alpha}^{} = 73.2~\text{MeV}$ is the binding momentum of 
\nuclide[5][\Lambda]{He}, and $a_{\Lambda\Lambda}^{}$ is the scattering length of the $S$-wave 
$\Lambda\Lambda$ scattering and its value is deduced from the data on the reaction of
$\nuclide[12]{C}(K^-,K^+\Lambda\Lambda X)$ in Ref.~\cite{GHH11}
as $a_{\Lambda\Lambda}^{} = -1.2\pm 0.6~\text{fm}$.
The coupling of the three-body contact interaction $g(\Lambda_c)$ is fixed by 
$B_{\Lambda\Lambda}=6.93~\text{MeV}$ as a function of the sharp cutoff $\Lambda_c$ introduced in 
the coupled integral equations. 
Further details of the integral equations can be found in Ref.~\cite{AO14}.

As mentioned above, the limit cycle behavior of the system can be explored by studying the 
homogeneous part of the integral equation in the asymptotic limit. 
As shown in Eq.~(9) of Ref.~\cite{AO14}, when we exclude the three-body interaction in the asymptotic 
limit $p\sim l \gg k, E, 1/a_{\Lambda\Lambda}^{}, \gamma_{\Lambda \alpha}^{}$, we have
\begin{equation}
1 = C_1\, I_1(s) + C_2\, I_2(s) \, I_3(s) ,
\label{eq:I123}
\end{equation}
where
\begin{equation}
C_1 = \frac{1}{2\pi}\frac{m_\alpha^{}}{\mu_{\Lambda\alpha}^{}}
\sqrt{
\frac{\mu_{\Lambda(\Lambda\alpha)}^{}}{\mu_{\Lambda\alpha}^{}}},
\quad
C_2 = 
\frac{\sqrt{2 m_\Lambda^{} \mu_{\Lambda(\Lambda\alpha)}^{} 
\mu_{\alpha(\Lambda\Lambda)}^{}
}}{\pi^2 \mu_{\Lambda\alpha}^{3/2}} ,
\end{equation}
and $\mu_{\alpha(\Lambda\Lambda)}^{} = 2 m_\Lambda^{} m_\alpha^{} /
(2m_\Lambda^{} + m_\alpha^{})$, etc.
The functions $I_{1,2,3}(s)$ are obtained by the Mellin transformation and their explicit expressions can 
be found in Ref.~\cite{AO14}.
The imaginary solution $s=\pm i s_0^{}$ indicates the limit cycle solution 
and we found the solution as
\begin{equation}
s_0^{} = 1.0496 \cdots .
\label{eq;s0}
\end{equation}
This shows that the $\Lambda\Lambda\alpha$ system exhibits the limit cycle and thus the three-body 
contact interaction should be included at LO for renormalization.

\begin{figure}[t]
\centering
\sidecaption
\includegraphics[width=8cm,clip]{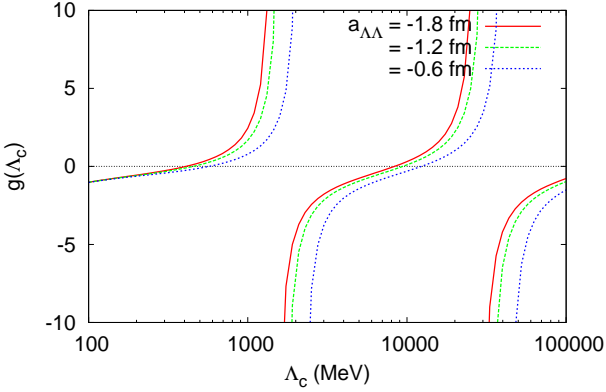}
\caption{
The coupling $g(\Lambda_c)$ as a function of $\Lambda_c$ for $a_{\Lambda\Lambda}^{}=-0.6$, 
$-1.2$, $-1.8$~fm, where the values of $g(\Lambda_c)$ are fitted by
$B_{\Lambda\Lambda}=6.93$~MeV of \nuclide[6][\Lambda\Lambda]{He}.
}
\label{fig-1}       
\end{figure}

In Fig.~\ref{fig-1}, we plot $g(\Lambda_c)$ as a function of $\Lambda_c$ with 
$a_{\Lambda\Lambda}^{} = -0.6$, $-1.2$, $-1.8$~fm.
One can verify that the curves clearly exhibit the limit cycle and the first divergence appears at 
$\Lambda_c \sim 1$~GeV.
In addition, a larger value of $|a_{\Lambda\Lambda}^{}|$ behaves as giving a larger attractive force 
and shifts the curves of $g(\Lambda_c)$ to the left as shown in Fig.~\ref{fig-1}.
Furthermore, the value of $s_0^{}$ can be estimated from the curves in Fig.~\ref{fig-1}.
The $(n+1)$-th values of $\Lambda_n$ at which $g(\Lambda_c)$ vanishes can be parameterized as 
$\Lambda_n = \Lambda_0\exp(n\pi/s_0^{})$ in the discrete scaling symmetry.
By using the second and third values of $\Lambda_n$ for the three values of $a_{\Lambda\Lambda}^{}$, 
we find $s_0^{} = \pi/\ln(\Lambda_2/\Lambda_1) \simeq 1.05$, which is in a very good agreement with 
the value of Eq.~(\ref{eq;s0}).

\begin{figure}[t]
\centering
\sidecaption
\includegraphics[width=8cm,clip]{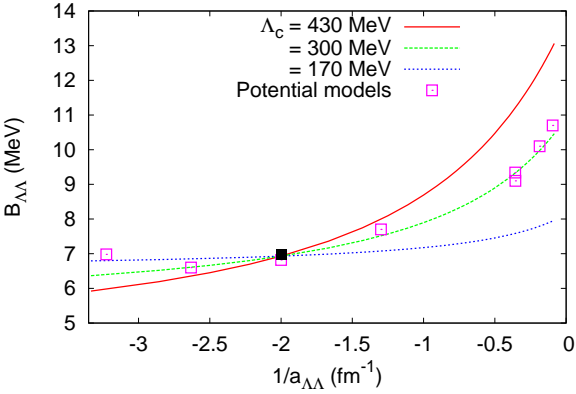}
\caption{
The two-$\Lambda$ separation energy $B_{\Lambda\Lambda}$ as a function
of $1/a_{\Lambda\Lambda}^{}$ for $\Lambda_c=170$, $300$, $430$~MeV, where
$g(\Lambda_c)$ is renormalized at the point of $B_{\Lambda\Lambda}=6.93$~MeV
and $1/a_{\Lambda\Lambda}^{}=-2.0$~fm$^{-1}$ that is marked by a filled square.
Open squares are the results from the potential models given in Table~5 of Ref.~\cite{FG02b}.
%
}
\label{fig-3}       
\end{figure}

Figure~\ref{fig-3} shows $B_{\Lambda\Lambda}$ as a function of 
$1/a_{\Lambda\Lambda}^{}$, while $g(\Lambda_c)$ is renormalized at the point marked by a filled square,
i.e., $B_{\Lambda\Lambda}=6.93$~MeV and $1/a_{\Lambda\Lambda}=-2.0$~fm$^{-1}$.
This leads to $g(\Lambda_c) \simeq -0.715$, $-0.447$, $-0.254$ for
$\Lambda_c=170$, $300$, $430$~MeV, respectively.
Open squares are the estimated values from the potential models given in Table 5 of Ref.~\cite{FG02b}.
The results show that the curves are sensitive to the cutoff value and the outputs from the potential 
models are remarkably well reproduced by the curve with $\Lambda_c=300$~MeV.

\section{Discussion and conclusions}

To summarize, the hypernucleus \nuclide[6][\Lambda\Lambda]{He} was studied in cluster EFT at LO
as a three-body $(\Lambda\Lambda\alpha)$ system.
The three-body contact interaction of the system exhibits the limit cycle, so it is needed to be promoted 
to LO so that the result is independent of the cutoff.
The equation to examine the limit cycle for \nuclide[6][\Lambda\Lambda]{He} is developed, which 
reproduces the numerical results for the limit cycle.
The two-$\Lambda$ separation energy $B_{\Lambda\Lambda}$ of \nuclide[6][\Lambda\Lambda]{He} can 
be well obtained even without introducing the three-body contact interaction, but it requires 
$\Lambda_c = 570 \sim 410~\text{MeV}$ for $a_{\Lambda\Lambda}^{} = -0.6 \sim -1.8~\text{fm}$. 
The cutoff values are comparable to the large scale of the theory, $\Lambda_H\simeq 170$~MeV.
This pattern is similar to that discovered already for the formation of three-body bound state without the 
three-body contact interaction such as the triton in pionless EFT~\cite{BHV99,AB10b}.
We also found that the dependence on $a_{\Lambda\Lambda}^{}$ is significant but the range of 
$a_{\Lambda\Lambda}^{}$ cannot be narrowed until we have more precise and ample 
experimental data for various physical quantities of the system.

It should also be noted that the equation we developed in Eq.~(\ref{eq:I123}) depends on the masses 
and the spin-isospin quantum numbers of the state but not on the details of dynamics.
It is well known that the imaginary solution of that equation implies the existence of the Efimov states in 
the unitary limit~\cite{BHV99}.
Even though our system is not close to the unitary limit, the presence of imaginary solution could imply 
the existence of a bound state.
Therefore, the integral equations in the asymptotic limit for three-body cluster systems would be useful to 
search for exotic states.%
\footnote{
This method is recently applied to study the existence of a putative $nn\Lambda$ bound system 
in Ref.~\cite{ARO15}.
}

In order to study the correlation between $B_{\Lambda\Lambda}$ and $a_{\Lambda\Lambda}^{}$
we introduce the three-body contact interaction. 
By varying the value of $\Lambda_c$, we found that the results of potential models can be remarkably 
well reproduced with $\Lambda_c =300$~MeV, which would demonstrate a nontrivial role of the two-pion 
exchange as a long range mechanism of the $\Lambda\Lambda$ interaction.
However, including two-pion exchanges in the present approach is highly nontrivial as it requires
to have $\Lambda_c > \Lambda_H \simeq 170~\text{MeV}$ and such a large cutoff demands the 
inclusion of various short range (or high momentum) degrees of freedom such as the first excitation
state of the $\alpha$. 
Investigating such mechanisms is beyond the scope of the present work but deserves further studies.

\begin{acknowledgement}
S.-I.A. is grateful to the organizers of the FB21 conference for hospitality and support.
He also acknowledges the support from the Sunmoon University Research Grant of 2015.
Y.O. was supported by the National Research Foundation (Grant No.\ NRF-2013R1A1A2A10007294) 
funded by the Ministry of Education of Korea. 
This work was also supported in part by the Ministry of Science, ICT, and Future Planning (MSIP)
and the National Research Foundation of Korea under Grant 
No.\ NRF-2013K1A3A7A06056592 (Center for Korean J-PARC Users).
\end{acknowledgement}

\end{document}